\documentclass[twocolumn]{aastex63}

\newcommand{\region}{G335.579--0.292}
\newcommand{\regshort}{G335--MM1}
\newcommand{\msun}{${\rm M}_\sun$}
\newcommand{\lsun}{${\rm L}_\sun$}
\newcommand{\kms}{km\,s$^{-1}$}

\received{XXX, 2020}
\revised{YYY, 2020}
\accepted{\today}
\submitjournal{ApJ}

\shorttitle{\regshort\ gas kinematics}
\shortauthors{Olguin et al.}

\graphicspath{{./}}

\begin{document}

\title{Digging into the Interior of Hot Cores with ALMA (DIHCA). \\I. Dissecting the High-mass Star-Forming Core G335.579--0.292 MM1}

\correspondingauthor{Fernando Olguin}
\email{folguin@phys.nthu.edu.tw}

\author[0000-0002-8250-6827]{Fernando A. Olguin} 
\affil{Institute of Astronomy and Department of physics, National Tsing Hua University, Hsinchu 30013, Taiwan} 

\author[0000-0002-7125-7685]{Patricio Sanhueza} 
\affiliation{National Astronomical Observatory of Japan, National Institutes of Natural Sciences, 2-21-1 Osawa, Mitaka, Tokyo 181-8588, Japan}
\affil{Department of Astronomical Science, SOKENDAI (The Graduate University for Advanced Studies), 2-21-1 Osawa, Mitaka, Tokyo 181-8588, Japan}

\author[0000-0003-0990-8990]{Andr\'es E. Guzm\'an}
\affiliation{National Astronomical Observatory of Japan, National Institutes of Natural Sciences, 2-21-1 Osawa, Mitaka, Tokyo 181-8588, Japan}

\author[0000-0003-2619-9305]{Xing Lu}
\affiliation{National Astronomical Observatory of Japan, National Institutes of Natural Sciences, 2-21-1 Osawa, Mitaka, Tokyo 181-8588, Japan}

\author{Kazuya Saigo}
\affiliation{National Astronomical Observatory of Japan, National Institutes of Natural Sciences, 2-21-1 Osawa, Mitaka, Tokyo 181-8588, Japan}

\author[0000-0003-2384-6589]{Qizhou Zhang}
\affiliation{Center for Astrophysics $|$ Harvard \& Smithsonian, 60 Garden Street, Cambridge, MA 02138, USA}

\author[0000-0001-9500-604X]{Andrea Silva}
\affiliation{National Astronomical Observatory of Japan, National Institutes of Natural Sciences, 2-21-1 Osawa, Mitaka, Tokyo 181-8588, Japan}

\author[0000-0002-9774-1846]{Huei-Ru Vivien Chen} 
\affil{Institute of Astronomy and Department of Physics, National Tsing Hua University, Hsinchu 30013, Taiwan} 

\author[0000-0003-1275-5251]{Shanghuo Li}
\affiliation{Korea Astronomy and Space Science Institute, 776 Daedeokdae-ro, Yuseong-gu, Daejeon 34055, Republic of Korea}

\author[0000-0002-9661-7958]{Satoshi Ohashi}
\affiliation{RIKEN Cluster for Pioneering Research, 2-1, Hirosawa, Wako-shi, Saitama 351-0198, Japan}

\author[0000-0001-5431-2294]{Fumitaka Nakamura}
\affiliation{National Astronomical Observatory of Japan, National Institutes of Natural Sciences, 2-21-1 Osawa, Mitaka, Tokyo 181-8588, Japan}
\affil{Department of Astronomical Science, SOKENDAI (The Graduate University for Advanced Studies), 2-21-1 Osawa, Mitaka, Tokyo 181-8588, Japan}

\author[0000-0003-4521-7492]{Takeshi Sakai}
\affiliation{Graduate School of Informatics and Engineering, The University of Electro-Communications, Chofu, Tokyo 182-8585, Japan.}

\author[0000-0003-3874-7030]{Benjamin Wu}
\affiliation{NVIDIA, 2788 San Tomas Expressway, Santa Clara, CA 95051, USA}
\affiliation{National Astronomical Observatory of Japan, National Institutes of Natural Sciences, 2-21-1 Osawa, Mitaka, Tokyo 181-8588, Japan}

\begin{abstract}
    We observed the high-mass star-forming region G335.579--0.292 with the Atacama Large Millimeter/submillimeter Array (ALMA) at 226\,GHz with an angular resolution of 0\farcs3 ($\sim$1000\,au resolution at the source distance).
    \region\ hosts one of the most massive cores in the Galaxy (G335--MM1). 
    The continuum emission shows that G335--MM1 fragments into at least five sources, while molecular line emission is detected in two of the continuum sources (ALMA1 and ALMA3).
    We found evidence of large and small scale infall in ALMA1 revealed by an inverse P-Cygni profile and the presence of a blue-shifted spot at the center of the first moment map of the CH$_3$CN emission.
    In addition, hot gas expansion in the innermost region is  unveiled by a red-shifted spot in the first moment map of HDCO and (CH$_3$)$_2$CO (both with $E_u > $ 1100 K). 
    Our modeling reveals that this expansion motion originates close to the central source, likely due to reversal of the accretion flow induced by the expansion of the \ion{H}{2} region, while infall and rotation motions originate in the outer regions.
    ALMA3 shows clear signs of rotation, with a rotation axis inclination with respect to the line of sight close to 90\degr, and a system mass (disk + star) in the range of 10--30\,\msun.
\end{abstract}


\section{Introduction}\label{sec:intro}

High-mass stars form in dense ``hot cores" (sizes of few 1000\,au), within massive molecular clumps that have been recently found to be preponderantly under global collapse \citep{2019ApJ...870....5J}. 
Observational evidence shows that many ``hot cores" are located in the intersection of large pc-scale filaments, the so-called hubs, which are fed through filaments \citep[e.g.,][]{2014ApJ...790...84L,2018ApJ...855....9L,2019A&A...629A..81T,2019ApJ...875...24C}. 
Similarly, at smaller scales ($<$0.1\,pc) streams of gas feeding individual cores have been observed \citep[e.g.,][]{2018MNRAS.478.2505I,2019A&A...628A...6S}.
At even smaller scales, spiral arms-like structures have been identified in disks within cores \citep[e.g.,][]{2020A&A...634L..11J}, which may play a role in the formation of multiple systems \citep[e.g.,][]{2010ApJ...708.1585K,2019A&A...632A..50A}.
Hence, a characterization of the kinematics of the gas at core and smaller scales is key to understanding how high-mass stars in single or multiple systems form, in addition to providing better constraints for numerical simulations.
In particular, cores inside infrared dark clouds \citep[IRDCs;][]{2006ApJ...641..389R,2009ApJS..181..360C,2012ApJ...756...60S,2019ApJ...886..102S,2019ApJ...886..130L,2020ApJ...896..110L} are more likely to be young enough that ionization plays a minor role, thus limiting the number of physical processes to account for in a detailed study.

The IRDC \region\ is a massive cloud with a mass of $5.5\times10^3$\,\msun\ at a distance of  3.25\,kpc \citep{2013A&A...555A.112P}.
\citet{2013A&A...555A.112P} identified two cores using 3\,mm Atacama Large Millimeter/submillimeter Array (ALMA) observations which are massive enough to form high-mass stars. 
The most massive core, \regshort, is one of the most massive cores in the Galaxy with a mass of 545\,\msun\ \citep[comparable to only few other cores, e.g.,][]{2015ApJ...802....6S}. 
\regshort\ is being fed by large scale infalling gas at a rate of ${\sim}10^{-3}$\,\msun\,yr$^{-1}$ \citep{2013A&A...555A.112P}.
\citet{2015A&A...577A..30A} estimated a bolometric luminosity of $1.6-1.8\times10^4$\,\lsun\ for \regshort.
Their radio cm wavelength observations revealed that \regshort\ further fragments into at least two sources (MM1a and MM1b) with  spectral indices consistent with hyper-compact (HC) \ion{H}{2} regions, and with free-free emission equivalent to those of zero age main sequence (ZAMS) B-type stars (9--10\,\msun).
Class II methanol, which are exclusively associated with high-mass star formation, and water masers have also been observed towards both regions within \regshort\  \citep{2010MNRAS.406.1487B,2011MNRAS.417.1964C,2015A&A...577A..30A}.
Regardless of these efforts, there are no studies so far of the gas kinematics at the size scales directly related to the star formation process ($\sim$1000 au).

In order to further characterize the formation of high-mass stars within this region, we have observed \regshort\ with ALMA at 226\,GHz to study the fragmentation process in this source and the kinematics of the gas at much smaller scales than previously reported. The observations of \regshort\ are part of a larger survey called Digging into the Interior of Hot Cores with ALMA (DIHCA), in which we have observed 30 high-mass star-forming regions. Details on the whole sample of DIHCA will be presented in a forthcoming paper (K. Ishihara et al. 2020, in preparation). 
In \S\ref{sec:obs} we present our ALMA observations of this source.
The results are presented in \S\ref{sec:res}.
Finally, our discussion and conclusions are presented in \S\ref{sec:dis} and \S\ref{sec:conc}.

\section{Observations}
\label{sec:obs}

We observed \regshort\ in band 6 (226.2\,GHz, 1.33\,mm) with ALMA during November 2016, cycle 4 (Project ID: 2016.1.01036.S; PI: Sanhueza). 
Observations were performed with the 12 m array using 41 antennas in a configuration similar to C40-5, with minimum and maximum baselines of 18.6 and 1100\,m, respectively.
With this configuration, the observations achieved a resolution of ${\sim}0\farcs3$ (${\sim}1000$\,au) and a maximum recoverable scale of $3\farcs2$ (${\sim}10000$\,au).
The spectral setup was divided in four spectral windows with a spectral resolution of 976.6\,kHz (${\sim}1.3$\,km\,s$^{-1}$) and a bandwidth of 1.875\,GHz.
These windows covered the frequency ranges between 233.5--235.5\,GHz, 231.0--233.0\,GHz, 216.9--218.7\,GHz and 219.0--221.0\,GHz.

\begin{figure}
\begin{center}
\includegraphics[angle=0,scale=0.59]{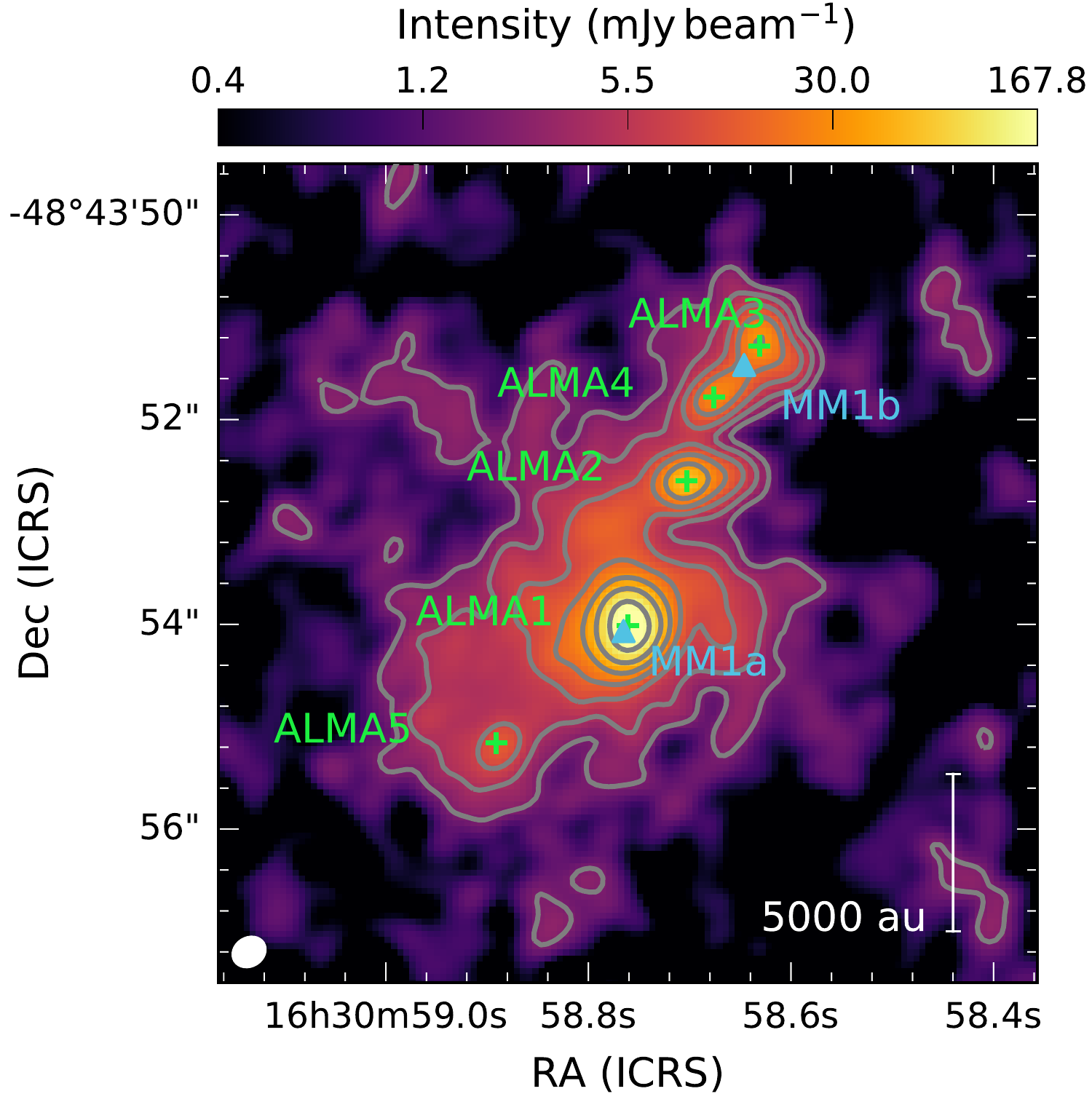}
\end{center}
\caption{ALMA continuum map of \regshort\ at 1.3\,mm in color scale and contours.  
The grey contours levels are $5, 10, 20, 40, 80, 160, 320\times\sigma_{\rm cont}$ with $\sigma_{\rm cont}=0.4$\,mJy\,beam$^{-1}$.
The green crosses and labels mark the peak position of the sources identified within this region and are labeled by brightness.
The position of the radio continuum sources from \citet{2015A&A...577A..30A} are marked with light blue triangles.
The beam size is shown in the lower left corner and corresponds to a scale of ${\sim}1000$\,au.
}
\label{fig:continuum}
\end{figure}

The data were calibrated using the CASA 4.7 reduction pipeline \citep[version r38377; ][]{2007ASPC..376..127M}.
According to the ALMA Proposer's User Guide, the estimated error in the absolute flux is 10\,\%. 
The data were then self-calibrated, and a continuum map from line-free channels and continuum subtracted data cubes were produced.
The channel identification procedure is detailed in Appendix~\ref{ap:contsub}.
The script and continuum subtraction pipeline are available on GitHub\footnote{GoContinuum: \url{https://github.com/folguinch/GoContinuum}.} under a MIT License. Version 2.0 of the script is archived in Zenodo \citep{olguin_fernando_2020_4302846}.
We used the {\sc tclean} task with Briggs weighting and robust parameter of 0.5 to image the data, resulting in a continuum sensitivity of 0.4\,mJy\,beam$^{-1}$ (vs. expected thermal noise level of 0.15\,mJy\,beam$^{-1}$). 
The cleaned continuum map is shown in Figure~\ref{fig:continuum}.
The synthesized beam of the continuum map is $0\farcs36\times0\farcs30$ with position angle (P.A.) of --58\degr. 
   
We used the automatic masking procedure YCLEAN \citep{2018ApJ...861...14C,2018zndo...1216881C} to CLEAN the data cubes for each spectral window.
YCLEAN iterates between CLEANing steps with increasingly deeper thresholds, and mask generating steps. 
The emission that is added to the masks and the CLEANing thresholds are based on the the image rms, residuals and beam secondary lobe at each step.
A noise level per channel between 4--6\,mJy\,beam$^{-1}$ was achieved.

\section{Results}
\label{sec:res}

In this section, we describe the \regshort\ high-mass star-forming region in 1.3\,mm dust continuum emission and selected molecular lines.
Those will be good tracers to investigate the morphology and kinematic structures of high-mass star forming cores.

\subsection{Morphology and Line Identification}

The single \regshort\ region identified at 3\,mm observations ($\sim$5\arcsec\ resolution) by \citet{2013A&A...555A.112P} fragments into at least five continuum sources (Figure~\ref{fig:continuum}).
These sources were identified by visual inspection.
The positions and 1.3\,mm fluxes of these sources were derived from a 2-D Gaussian fit to the continuum emission and are listed in Table~\ref{tab:props}.
The fit indicates that the sources are elongated with ratios between their axes in the 1.2--2.0 range.
The brightest of the five 1.3\,mm sources identified by us (ALMA1) matches with the position of one of the two radio continuum sources identified by \cite{2015A&A...577A..30A}, MM1a.
The second radio source, MM1b, is located in between of two of the 1.3\,mm sources (ALMA3 and ALMA4).

\begin{deluxetable*}{lcccccc}
\tablecaption{Continuum source properties\label{tab:props}}
\tablewidth{0pt}
\tablehead{
\colhead{ALMA} & \colhead{RA} & \colhead{Dec} & \colhead{$I_{\rm peak}$} & \colhead{$F_{\rm total}$} & \colhead{Deconvolved Size} & \colhead{$v_{\rm LSR}$}\\
\colhead{Source} & \colhead{$[h:m:s]$} & \colhead{$[\degr:\arcmin:\arcsec]$} & \colhead{(mJy\,beam$^{-1}$)} & \colhead{(mJy)} & \colhead{(\arcsec)} & \colhead{(km s$^{-1}$)}
}
\startdata
1 & 16:30:58.761 & -48:43:54.10 & 209.8 & $566\pm28$ & $0.50\times0.39$ & --46.9 \\ 
2 & 16:30:58.703 & -48:43:52.60 & 60.2  & $114\pm12$ & $0.47\times0.26$ & -- \\
3 & 16:30:58.631 & -48:43:51.28 & 34.1  & $ 88\pm8 $ & $0.51\times0.39$ & --47.6 \\
4 & 16:30:58.676 & -48:43:51.78 & 27.8  & $ 48\pm4 $ & $0.40\times0.21$ & -- \\
5 & 16:30:58.889 & -48:43:55.18 & 12.6  & $ 48\pm4 $ & $0.69\times0.51$ & -- \\
\enddata
\tablecomments{Positions in ICRS coordinate standard.}
\end{deluxetable*}

We have identified two sources potentially associated with MM1b (ALMA3 and ALMA4).
\citet{2015A&A...577A..30A} argue that the spectral indices from radio emission for MM1b are consistent with a collimated jet or a compact \ion{H}{2} region.
They suggested that this emission is not associated with MM1a, but rather with another source due to the presence of CH$_3$OH maser emission closer to MM1b. 
They discarded the jet hypothesis due to the misalignment of MM1b with the HNC emission tracing the molecular outflow associated with MM1a.
Given the offset position of MM1b to the ALMA sources, we argue that the radio emission is possibly tracing the lobe of an ionized jet likely associated with the high-mass source ALMA3.
However, higher resolution observations at wavelengths $\lambda\geq3$\,mm are needed to resolve the radio emission and determine its orientation with respect to the ALMA sources.

Line emission is detected mainly towards ALMA1 and ALMA3.
Multiple lines are detected towards ALMA1, which is characteristic of young high-mass sources, like hot cores and ultra-compact \ion{H}{2} regions \citep[e.g.,][]{1998A&AS..133...29H}.
A lower number of species is detected towards ALMA3. 
The lines detected in ALMA3 are weaker than those of ALMA1.
We focus our analysis in a handful of lines to explore the kinematics of these two cores.
These lines and their properties are summarized in Table~\ref{tab:lines}.
The molecular lines detected in other sources correspond to transitions tracing more extended material (e.g., $^{13}$CO, C$^{18}$O), and thus cannot be assigned to a source in particular.
The lack of line emission towards the other sources may indicate that they are in an earlier evolutionary stage, likely in the prestellar phase.

\begin{deluxetable}{llccc}
\tablecaption{Summary of lines analyzed\label{tab:lines}}
\tablewidth{0pt}
\tablehead{
\colhead{Molecule} & \colhead{Transition} & \colhead{Frequency} & \colhead{$E_{u}$} & Ref.\\
\colhead{} & \colhead{} & \colhead{(GHz)} & \colhead{(K)} & \colhead{}
} 
\startdata
CH$_3$CN           & $J_K=12_4-11_4$               & 220.6792869 & 183  & (1)  \\
CH$_3$CN           & $J_K=12_7-11_7$               & 220.5393235 & 418  & (1)  \\
CH$_3$CN           & $J_K=12_8-11_8$               & 220.4758072 & 526  & (1)  \\
$^{13}$CO          & $J=2-1$                     & 220.3986842 & 16   & (2) \\
SiO                & $J=5-4$                     & 217.1049800 & 31   & (2) \\
H$_2$CO            & $J_{K_a,K_c}=3_{0,3}-2_{0,2}$         & 218.2221920 & 21   & (2) \\
HDCO               & $J_{K_a,K_c}=28_{5,23}-29_{3,26}$     & 220.0294078 & 1460 & (2) \\
$({\rm CH}_3)_2$CO & $J_{K_a,K_c}=54_{27,28}-54_{26,29}$ AE & 231.6868319 & 1148 & (1) \\
\enddata
\tablerefs{(1) Jet Propulsion Laboratory \citep[JPL,][]{1998JQSRT..60..883P};
(2) Cologne Database for Molecular Spectroscopy \citep[CDMS,][]{2005JMoSt.742..215M}.}
\end{deluxetable}

\begin{figure}
\epsscale{1.18}
\plotone{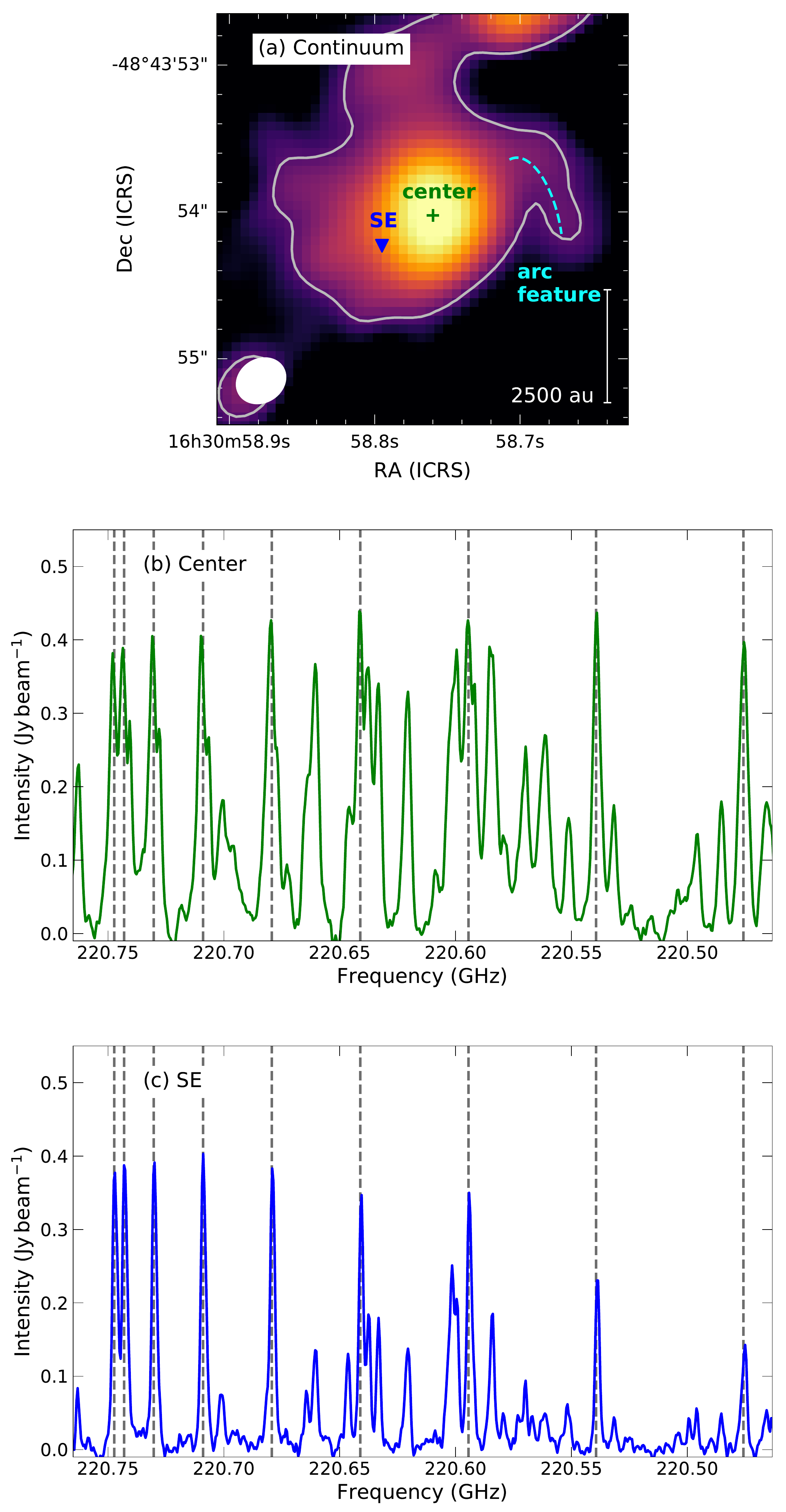}
\caption{G335 ALMA1 continuum map and CH$_3$CN $J=12-11\,K=0$ to $8$ line emission examples.  
(a) Locations of the example spectra marked over the continuum map. 
The light blue dashed line shows the arc shaped structure.
Contour level corresponds to $20\times\sigma_{\rm cont}$ with $\sigma_{\rm cont}=0.4$\,mJy\,beam$^{-1}$.
(b) Line emission towards continuum peak, which is marked with a green cross on (a).
(c) Line emission towards the south-east of the dust peak position, in a less dense region marked with a blue triangle in (a).
Dashed gray vertical lines mark the rest frequency of the $K$ transitions (0--8 from left to right) corrected by the systemic velocity ($v_{\rm LSR}=-46.9$\,km\,s$^{-1}$).
\label{fig:ch3cn:kladder}}
\end{figure}

We estimated the local standard of rest (LSR) velocity ($v_{\rm LSR}$) of ALMA1 from the CH$_3$CN $J=12-11$ $K=7$ and $8$ transitions.
For ALMA3, we estimated the $v_{\rm LSR}$ from the centroid of the $K=7$ line, which is equivalent to the velocity at the position of the source from its first moment map.
The velocities are listed in Table~\ref{tab:props}.
Figure~\ref{fig:ch3cn:kladder} shows the continuum of ALMA1 and example CH$_3$CN spectra from two locations marked on the continuum image. 
The saturated CH$_3$CN line emission in Figure~\ref{fig:ch3cn:kladder}b indicates that the lower $K$ transitions have become optically thick towards the center position of ALMA1.
These transitions even show self-absorbed profiles, except for $K=7$ and $8$, which are optically thinner. 
The spectrum shown toward the south-east of the peak position of ALMA1 (Figure~\ref{fig:ch3cn:kladder}c) is typical of optically thin spectra around the central region, displaying Gaussian-like profiles and not self-absorption features. 
Although the gas traced by high $K$-ladder transitions is hotter than low $K$ transitions and may thus correspond to a different velocity component, the derived LSR velocity is consistent with other species and previous estimations of the source $v_{\rm LSR}$ at larger scales \citep[$v_{\rm LSR}=-46.6$\,km\,s$^{-1}$;][]{1996A&AS..115...81B}. 
To determine the $v_{\rm LSR}$ of ALMA1, the lines were fitted with a Gaussian profile and the adopted velocity correspond to the average $v_{\rm LSR}$ of both $K=7$ and 8 transitions.

\subsection{Physical Properties}
\label{sec:res:physprops}

From the 2-D Gaussian fit to the dust continuum image, we estimate the size of the sources.
The radius is defined as the geometric mean of the deconvolved semi-major and minor axes (see Table~\ref{tab:props}).
The results are listed in Table~\ref{tab:physprops}.
All sources are compact (${\sim}400-800$\,au) with radii typical of inner envelope/disc scales ($<5000$\,au).

\begin{deluxetable}{lccccc}
\tablecaption{Source physical properties\label{tab:physprops}}
\tablewidth{0pt}
\tablehead{
\colhead{ALMA} & \colhead{$R$} & \colhead{$T$} & \colhead{$M_d$} & \colhead{$N_{{\rm H}_2}$} & \colhead{$N_{\rm CH_3CN}$}\\
\colhead{Source} & \colhead{(au)} & \colhead{(K)} & \colhead{(\msun)} & \colhead{($10^{24}$\,cm$^{-2}$)} & \colhead{(cm$^{-2}$)}
}
\startdata
1$^{a}$ & 710 & 100      & 19      & 10.4 & -- \\
        &     & 300      & 6.2     & 3.3  & -- \\ 
2$^{b}$ & 570 & ${>}20$  & ${<}24$ & 18.7 & -- \\
3       & 730 & 290      & 1.0     & 0.6  & $10^{13}$ \\
4$^{b}$ & 470 & ${>}20$  & ${<}10$ & 8.7  & -- \\
5$^{b}$ & 960 & ${>}20$  & ${<}10$ & 3.9  & -- \\
\enddata
\tablecomments{The radius, $R$, corresponds to half of the geometric mean of the deconvolved sizes (FWHM) from Table~\ref{tab:props}.
\\
$^{a}$ The dust temperatures are an approximation (see Section~\ref{sec:res:physprops}).\\
$^{b}$ CH$_3$CN emission is not detected towards these sources, hence the temperature is a lower limit, and the gas masses and column densities are upper limits.}
\end{deluxetable}

To estimate the dust temperature, we attempt to fit the CH$_3$CN $J=12-11$ $K=0$ to $7$ transitions towards the continuum peak of the sources with local thermal equilibrium (LTE) 1-D model.
However, the line emission towards ALMA1 becomes quickly saturated, hence we only fit the temperature towards ALMA3.
The fit was performed with the Markov Chain Monte Carlo method within XCLASS \citep{2017A&A...598A...7M}.
We obtain a gas temperature at the continuum peak of ALMA3 of 290\,K.
For ALMA1, we estimate a dust mass using a temperature of 100\,K and 300\,K.
The former correspond to roughly the brightness temperature at which the lower $K$ transition lines saturate.
As the emission is becoming optically thick, the brightness temperature should converge to the kinetic temperature.
However, the beam filling factor is likely less than one \citep[e.g.,][]{2009ApJ...704L...5S}, hence this provides a temperature lower limit and in turn a dust mass upper limit.
The latter temperature is based on the result for ALMA3.

Assuming that the dust and gas temperatures are in equilibrium, we estimate the gas mass as
\begin{equation}\label{eq:dustmass}
    M_d = \frac{F_\nu d^2 R_{gd}}{\kappa_\nu B_\nu(T_d)}
\end{equation}
with $F_{1.3{\rm mm}}$ the flux density from Table~\ref{tab:props}, $d=3.25$\,kpc the source distance, the gas-to-dust mass ratio $R_{\rm gd}=100$, the dust opacity $\kappa_{1.3{\rm mm}}=1$\,cm$^2$\,gr$^{-1}$ \citep{1994A&A...291..943O}, and $B_\nu$ the Planck blackbody function.
The H$_2$ column density is estimated from the dust emission as
\begin{equation}
  N_{{\rm H}_2} = \frac{I_\nu R_{gd}}{B_\nu(T_d) \kappa_\nu \mu_{\rm H_2} m_{\rm H}}
\end{equation}
with $I_{1.3{\rm mm}}$ the peak intensity from Table~\ref{tab:props}, $\mu_{\rm H_2}=2.8$ the molecular weight per hydrogen molecule \citep[e.g.,][]{2008A&A...487..993K}, and $m_{\rm H}$ the atomic hydrogen mass.
The results are listed in Table~\ref{tab:physprops}.
We estimate that the contribution from free-free emission to the 1.3\,mm flux densities is less than 5\,mJy for ALMA1 and less than 1\,mJy for ALMA3 from the fit to the radio cm observations in \citet[][]{2015A&A...577A..30A}, and are thus negligible.

The 1.3\,mm flux density of \regshort\ is 1.4\,Jy from aperture photometry and 0.9\,Jy by summing the values in Table~\ref{tab:props}.
The expected flux at 1.3\,mm from the spectral energy distribution in \citet[][]{2015A&A...577A..30A}, obtained by interpolating the data points at 870\,\micron\ and 3.2\,mm, is ${\sim}2.1$\,Jy.
On the other hand, the sum of the masses is ${\sim}10$\% of the total mass estimated by \citet{2013A&A...555A.112P} from 3.2\,mm observations.
The discrepancy in mass can be explained primarily by the temperature estimates, followed by extended emission not included in the 2-D Gaussian fit measurements, and finally emission filtered out by the interferometer. 

The mass of the gas reservoir of ALMA3 is 1\,\msun\ (Table~\ref{tab:physprops}).
However this value is likely a lower limit, because cooler regions of the envelope also contribute to the dust emission.
Emission from the region may have also been filtered out by the interferometer.

\subsection{Kinematics}
\label{sec:res:kin}

\begin{figure}
\plotone{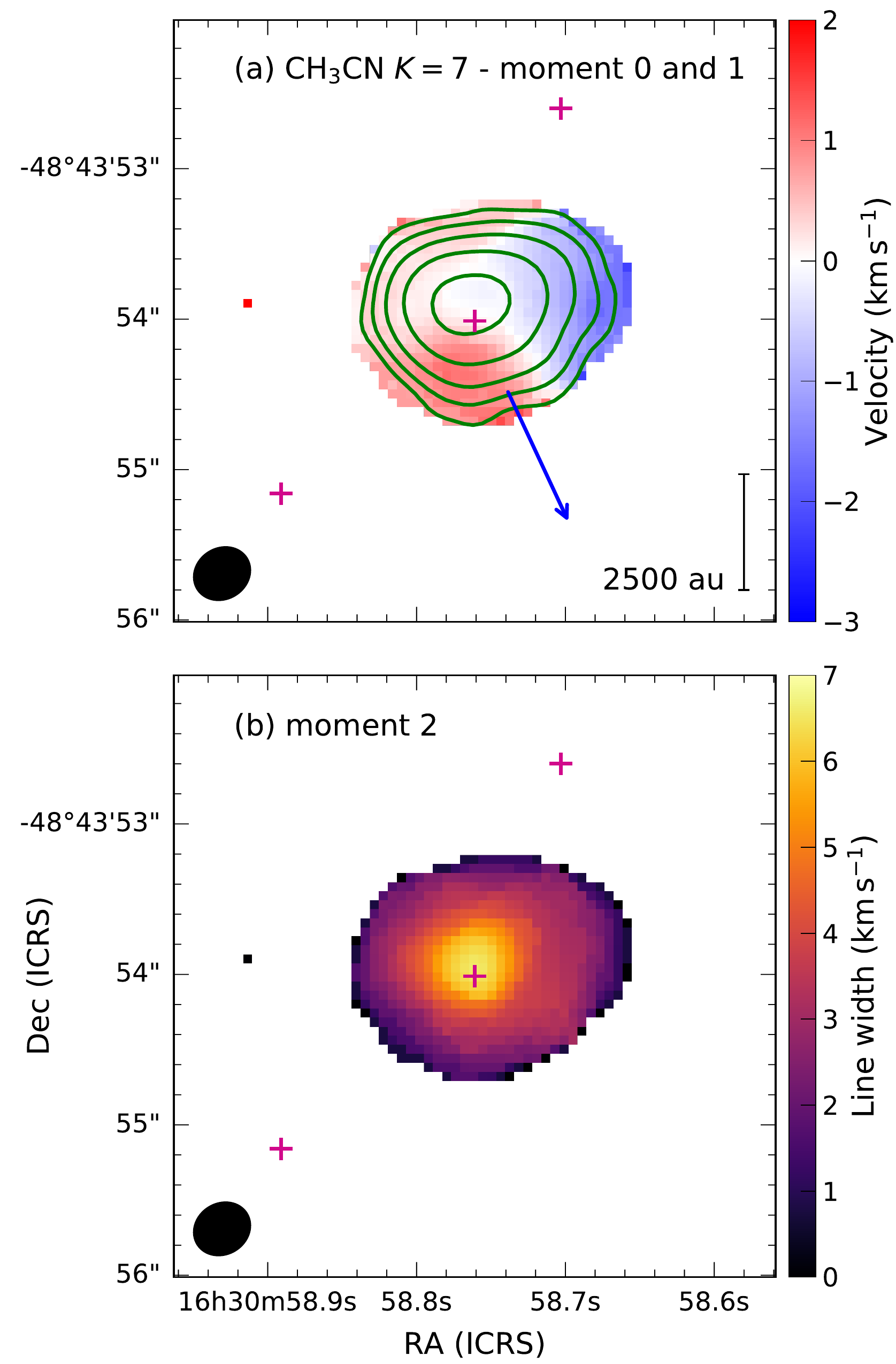}
\caption{G335 ALMA1 moment maps from CH$_3$CN $J=12-11$ $K=7$ line emission. 
(a) First moment map with zeroth moment shown in green contours.
Zero systemic velocity corresponds to the $v_{\rm LSR}$ (--46.9\,\kms).
Contour levels are 5, 10, 20, 40, $80\times \sigma_{\rm rms}$ with $\sigma_{\rm rms}=31$\,mJy\,beam$^{-1}$\,km\,s$^{-1}$.
The blue arrow shows the direction of the blue-shifted outflow lobe (see Section~\ref{sec:res:kin}).
(b) Second moment map.
The pink crosses mark the position of the continuum sources, and the beam size is shown in the lower left corners.
The physical scale is shown in the lower right corner of panel (a).
\label{fig:ch3cn:k7a}}
\end{figure}

\begin{figure}
\plotone{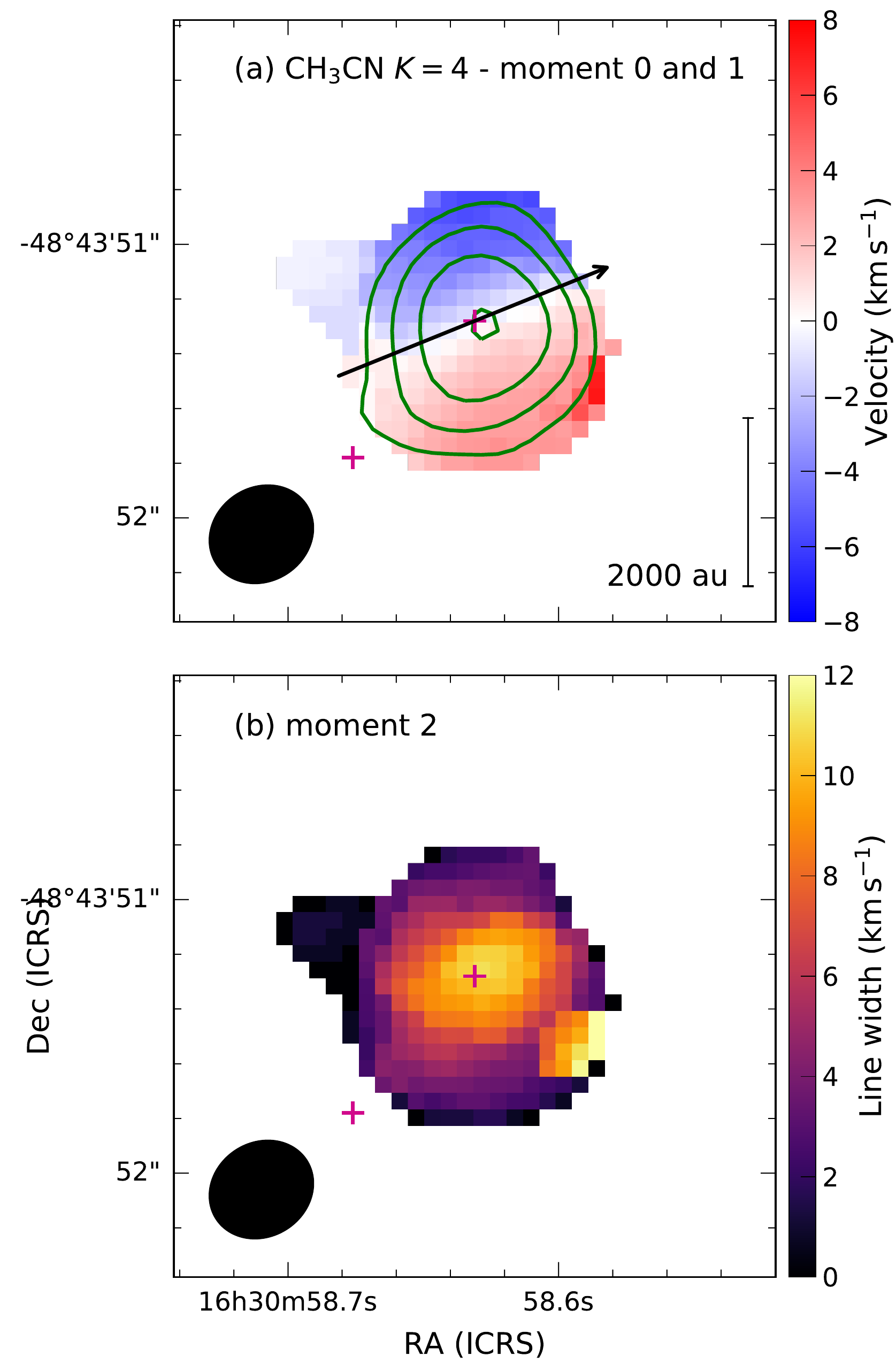}
\caption{G335 ALMA3 moment maps of CH$_3$CN $J=12-11$ $K=4$ line emission. 
(a) First moment map with zeroth moment shown in green contours.
Zero systemic velocity corresponds to the source $v_{\rm LSR}$ (--47.6\,\kms).
Contour levels are 5, 10, 20, $40\times \sigma_{\rm rms}$ with $\sigma_{\rm rms}=51$\,mJy\,beam$^{-1}$\,km\,s$^{-1}$.
The arrow shows the derived direction of the rotation axis (following the right hand rule; see Section~\ref{sec:res:kin}).
(b) Second moment map.
The pink crosses mark the position of the continuum sources, and the beam size is shown in the lower left corners.
The physical scale is shown in the lower right corner of panel (a).
\label{fig:ch3cn:k4b}}
\end{figure}

Line emission from CH$_3$CN transitions, a commonly used tracer of gas  rotation, is detected only towards two sources (ALMA1 and ALMA3). 
In ALMA1, transitions $K=0$ to $5$ are saturated and blended with other molecular lines (e.g., Figure~\ref{fig:ch3cn:kladder}).
We therefore calculated the moments 0, 1, and 2 for transition $K=7$, which are presented in Figure~\ref{fig:ch3cn:k7a} (line width is displayed instead of velocity dispersion).
For ALMA3, the transition $K=4$ is used in the moments shown in Figure~\ref{fig:ch3cn:k4b} as there is less contamination from other molecular lines.
First and second moments were calculated only with data over $5\sigma_{\rm rms}$ with $\sigma_{\rm rms}=5$\,mJy\,beam$^{-1}$.
Line emission from CH$_3$CN is only detected in ALMA1 and ALMA3 as shown by the contours in Figures~\ref{fig:ch3cn:k7a}a and \ref{fig:ch3cn:k4b}a.
The deconvolved size of the emission in the contours of Figure~\ref{fig:ch3cn:k7a}a is 0\farcs6 ($\sim2000$\,au), thus it is likely tracing the inner region of the circumstellar envelope and/or disk.

\begin{figure}
\epsscale{1.1}
\plotone{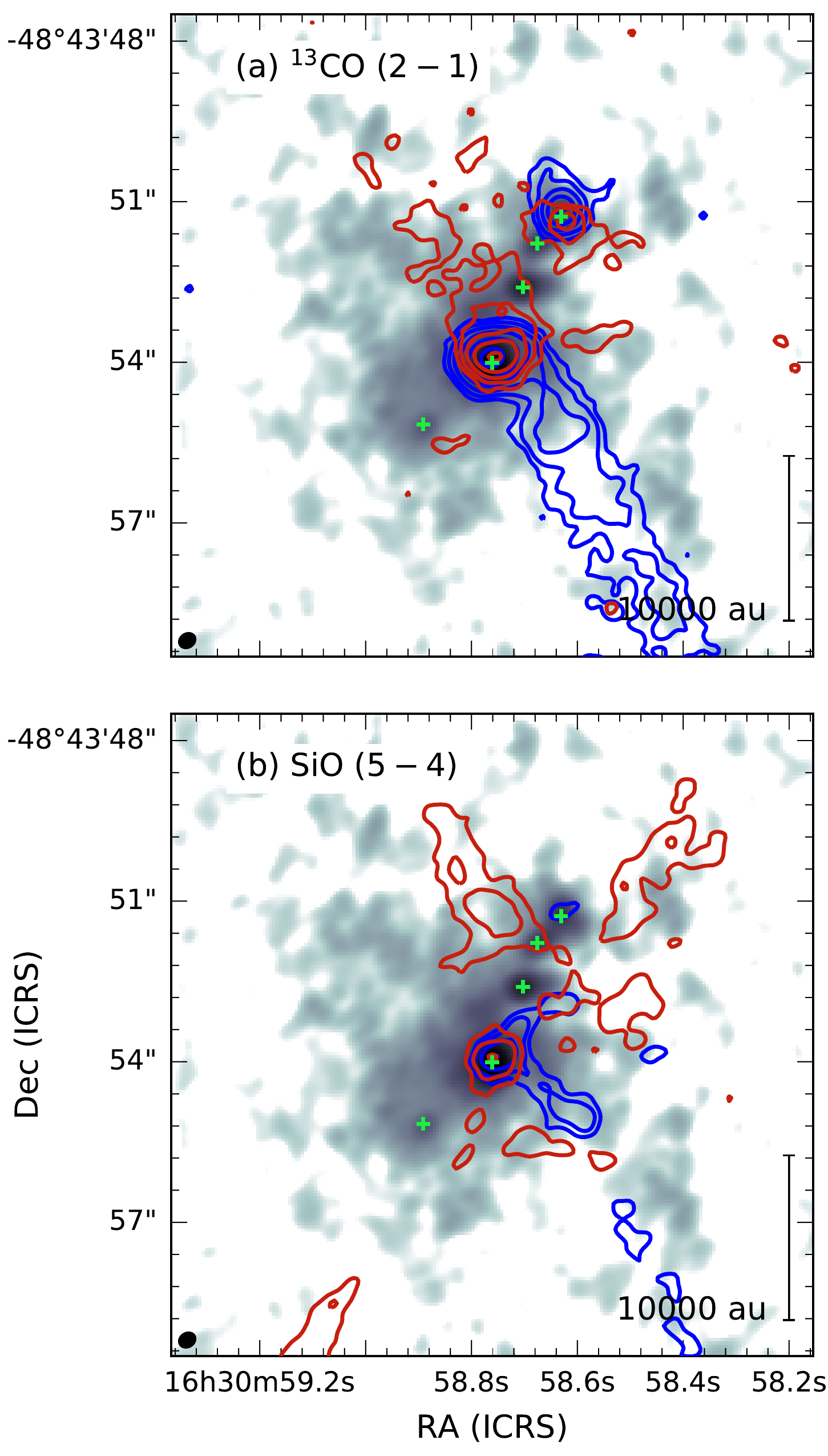}
\caption{Blue- and red-shifted zeroth order moment from $^{13}$CO $J=2-1$ and SiO $J=5-4$.
Blue and red contours show the blue- and red-shifted emission,  respectively.
Contour levels are 3, 6, 12, 24, 48 and 96$\times \sigma_{\rm rms}$ with $\sigma_{\rm rms}=47$\,mJy\,beam$^{-1}$\,km\,s$^{-1}$ for $^{13}$CO and 38\,mJy\,beam$^{-1}$\,km\,s$^{-1}$ for SiO.
Gray scale map corresponds to the continuum emission and green crosses mark the position of the ALMA sources.
The beam is shown in the lower left corner. 
\label{fig:13co:sio}}
\end{figure}

Outflow emission is detected in the $^{13}$CO $J=2-1$ and SiO $J=5-4$ transition lines. 
We separated the red- and blue-shifted line components to study the directions of the flows (Figure~\ref{fig:13co:sio}).
The blue and red windows are separated $\pm3.25$\,\kms\ from the $v_{\rm LSR}$ and have widths of $\sim22.75$ and $13$\,\kms\ for $^{13}$CO and SiO, respectively.
A clear molecular flow is detected towards ALMA1 in the NE-SW direction.
From the blue-shifted emission, we estimate a position angle ${\rm PA}\sim210\degr$.
Additionally, the arc shaped structure observed in the dust emission towards the west of ALMA1 (Figure~\ref{fig:ch3cn:kladder}a) is likely associated with emission from the base of the outflow cavity as seen in $^{13}$CO.
This may be the result of a wide angle wind interacting with the envelope \citep[e.g.,][]{2015ApJ...800...86K}, as observed in the diffuse dust emission and outflow emission of G16.64+0.16 \citep{2018A&A...620A..31M}.
Towards ALMA3, $^{13}$CO seems to be tracing the envelope of the source with the blue- and red-shifted orientation consistent with the rotation pattern observed in CH$_3$CN.
The red-shifted SiO emission towards the NW shows structures that may be associated with ALMA3 (SE-NW direction, ${\rm P.A.}{\sim}-45\degr$) and ALMA4 (SW-NE direction, ${\rm P.A.}{\sim}45\degr$).
However, since ALMA3 seems to be closer to edge-on given the symmetry of the CH$_3$CN first moment map and the origin of the SiO emission does not coincide with ALMA3, we cannot rule out that the SiO emission is related to an unresolved source.

\begin{figure*}
\epsscale{1.2}
\plotone{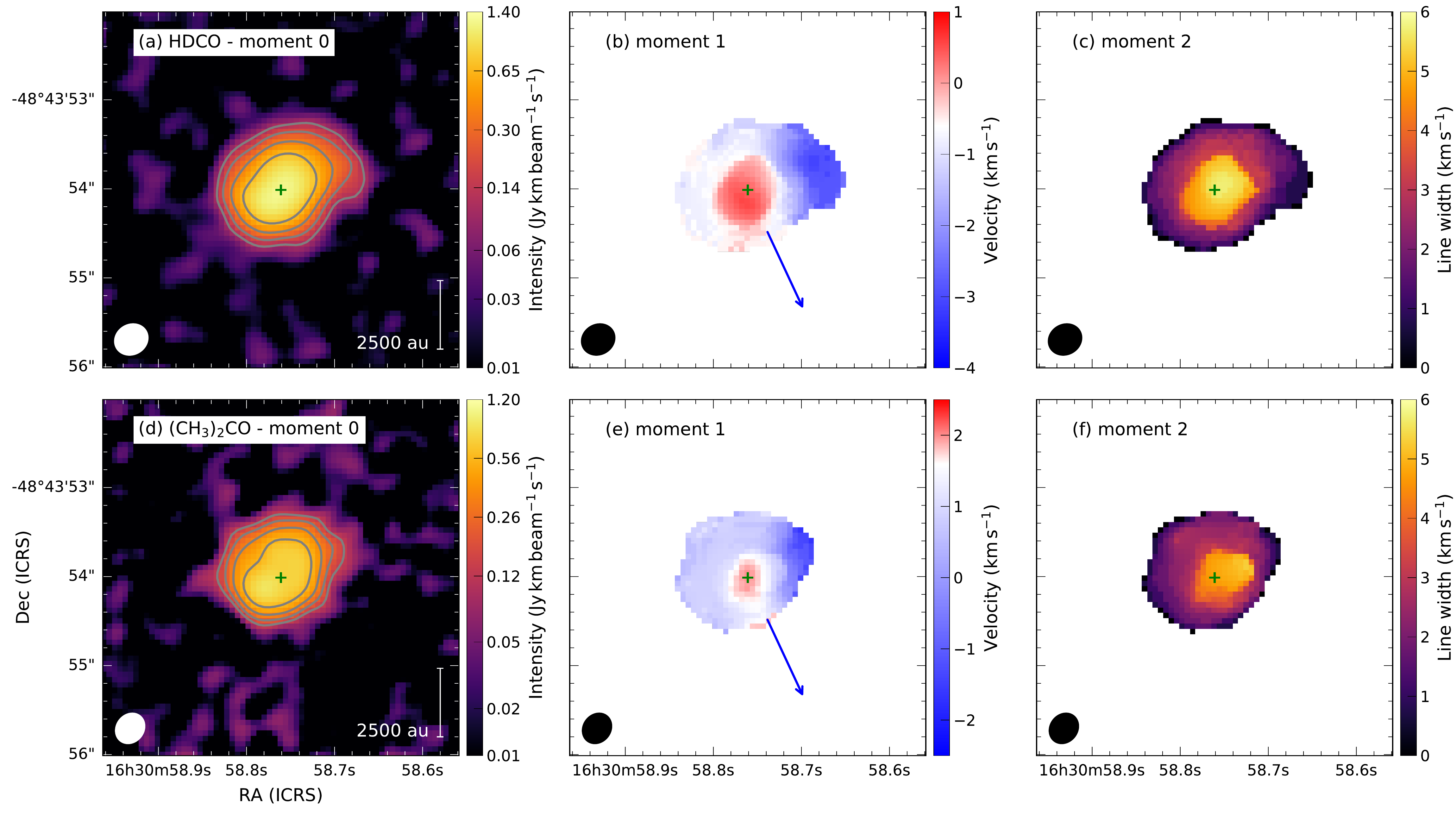}
\caption{Moment maps of hot lines ($E_{u}>1100$\,K) emission from G335--MM1 ALMA1. 
(a)--(c) Moments 0, 1 and 2 from the HDCO $(28_{5,23} - 29_{3,26})$ transition ($E_{u}=1460$\,K). 
(d)--(f) Same as upper row but for (CH$_3$)$_2$CO $(54_{27,28}-54_{26,29})$\,AE transition ($E_{u}=1148$\,K).
Zero systemic velocity in (b) and (e) corresponds to the source LSR velocity, $v_{\rm LSR}=-46.9$\,\kms, and the color scales are shifted to highlight the velocity structure around the continuum source.
Blue arrows in (b) and (e) show the direction of the blue-shifted outflow.
\label{fig:hotlines:moments}}
\end{figure*}

We have identified two lines with upper energy levels higher than 1000\,K: HDCO $(28_{5,23} - 29_{3,26})$ with $E_{\rm up}$ equal to 1460\,K and (CH$_3$)$_2$CO $(54_{27,28}-54_{26,29})$\,AE with $E_{\rm up}$ equal to 1140\,K. 
Their moment maps are displayed in Figure~\ref{fig:hotlines:moments}. 
The emission of these two lines is compact, but resolved towards ALMA1. 
Both lines are likely tracing the innermost regions of the hot core given the high temperatures required to be excited.
The first moment maps of these lines show a consistent velocity gradient with that of CH$_3$CN, but remarkably, the lines are red-shifted rather than blue-shifted towards the center of the source.

\section{Discussion}\label{sec:dis}

Here, we discuss the results focused primarily on the kinematics of ALMA1 and ALMA3.

\subsection{ALMA1}\label{sec:dis:alma1}

\subsubsection{Large Scale Infall}

The velocity field in ALMA1 presents an overall gradient from east to west (Figure~\ref{fig:ch3cn:k7a}a).
However this gradient is likely masked as a result of a combination of gas motions.
Possible reasons that make velocity gradients difficult to identify are \citep[e.g.,][]{Silva17}: 
(i) CH$_3$CN is possibly tracing outflowing motions, for instance, from gas removed from a disk surface by stellar winds, together with rotation and infall \citep[e.g.,][]{2017A&A...603A..10B} making the interpretation of the velocity gradients less straightforward; 
(ii) the observed velocity field is potentially produced by the combination of gas motions due to the presence of unresolved sources; 
(iii) the orientation of the protostellar disk is close to face-on; 
and/or (iv) infall or expansion motions. 

The first moment map of the $K=7$ transition presented in Figure~\ref{fig:ch3cn:k7a}a shows a spot of velocities close to zero at the center of the core enclosed mainly by the $80{\times}\sigma_{\rm rms}$ level of the zeroth moment contour.
These velocities are bluer than expected for a smooth transition of velocity from east to west, assuming CH$_3$CN is tracing the rotation of the core envelope/disk.
Henceforth we refer to this region as the blue-shifted spot.
As shown in Figure~\ref{fig:ch3cn:kladder}, the CH$_3$CN lines up to $K=4$ show an absorption feature at the center of the line, resembling the blue-asymmetry characteristic of infall motions \citep[e.g.,][]{1998ApJ...494..636Z} as is the blue-shifted spot \citep[e.g.,][]{2019A&A...626A..84E}.
This blue-asymmetric profile is not evident in larger, likely optically thinner, $K$ transitions, but the effect of the collapse can still be detected through the first moment map.
The first moment map spot of blue-shifted velocities (with respect to the $v_{\rm LSR}$) is thus likely produced by line profiles that are blue-skewed.
To further investigate the infall hypothesis, we search for blue-shifted emission coupled with red-shifted absorption against the continuum source, also known as an ``inverse P-Cygni profile". 
Such features provide unambiguous evidence for infall toward the central protostar \citep[e.g.,][]{Evans15}.  

\begin{figure}
\epsscale{1.15}
\plotone{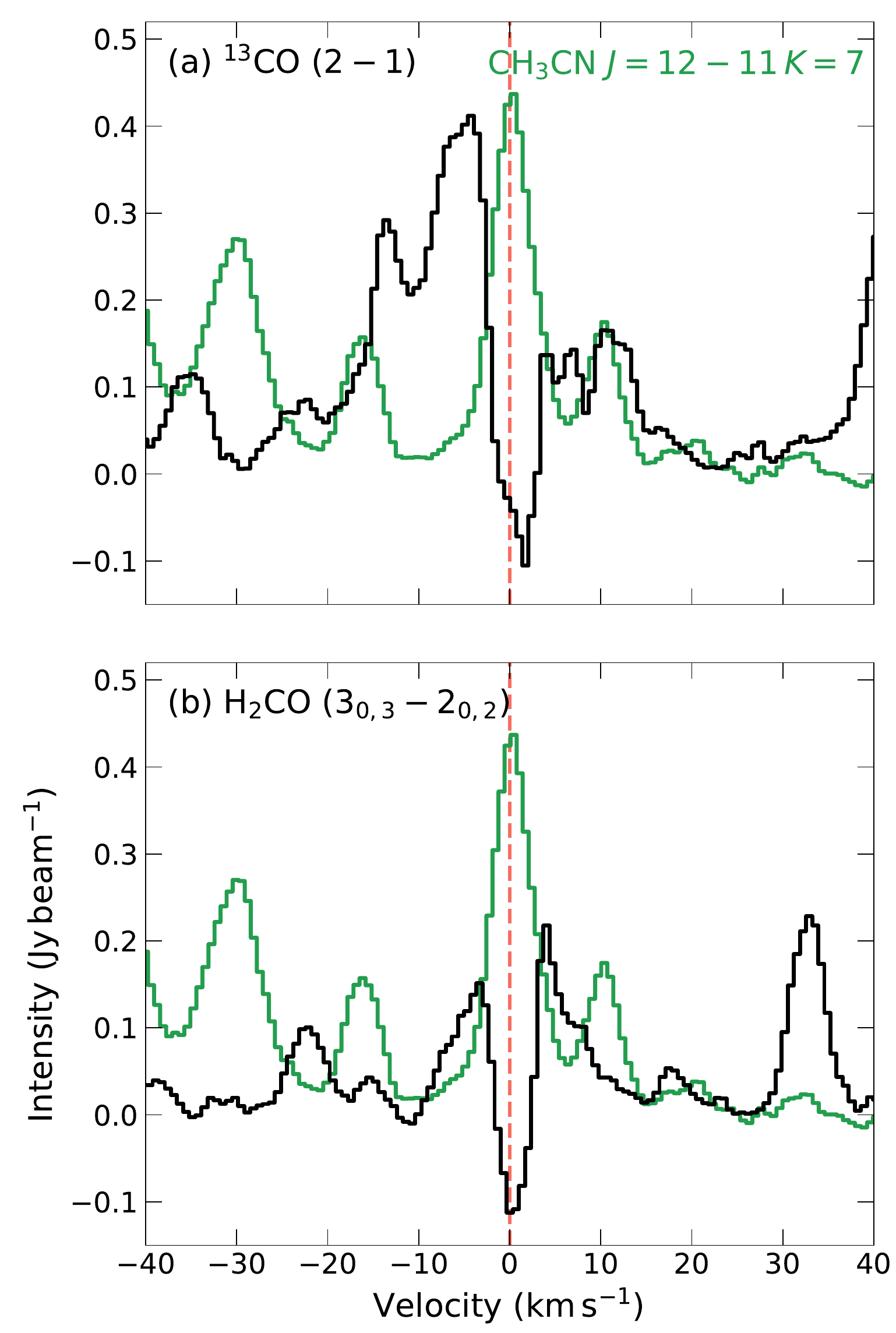}
\caption{Examples of absorption line profiles towards ALMA1. 
The black lines in (a) and (b) present the $^{13}$CO $(2-1)$ and H$_2$CO $(3_{0,3}-2_{0,2})$ line emission, respectively.
The green lines show the optically thin CH$_3$CN $J=12-11\,K=7$ transition.
\label{fig:13co:h2co}}
\end{figure}

\begin{figure}
\epsscale{1.15}
\plotone{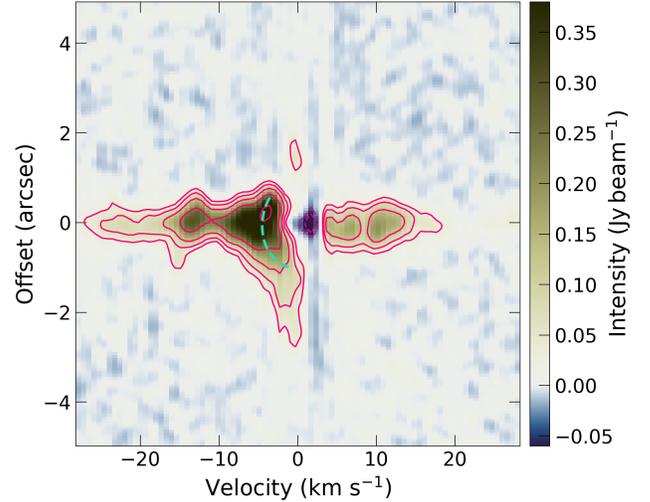}
\caption{Position velocity (pv) map of the $^{13}$CO line emission towards ALMA1.
The position angle of the pv cut is 90\degr, i.e. close to perpendicular with respect to the outflow emission in Figure~\ref{fig:13co:sio}.
The red contours correspond to $-9.5, -5.0, 5.0, 9.5, 18, 34, 65\times\sigma_{\rm rms}$ with $\sigma_{\rm rms}=6.7$\,mJy\,beam$^{-1}$.
The dashed green line highlights the ``C" shape produced by infalling motions.
\label{fig:13co:pvmap}}
\end{figure}

Among the myriad of lines included in the spectral setup, those that trace a more extended gas distribution display absorption against the continuum only at the center position of ALMA1 (and not anywhere else).
Figure~\ref{fig:13co:h2co} shows the spectrum of $^{13}$CO $(2 - 1)$ and H$_2$CO $(3_{0,3} - 2_{0,2})$ at the peak position of the dust continuum of ALMA1.
The $^{13}$CO displays an absorption against the continuum that is red-shifted with respect to the $v_{\rm LSR}$ of --46.9\,\kms\ by 1.7\,\kms.
On the other hand, the H$_2$CO line in Figure~\ref{fig:13co:h2co}b shows a dip consistent with the source $v_{\rm LSR}$. 
In order to be excited, $^{13}$CO requires lower densities and temperatures than H$_2$CO. 
We therefore suggest that the absorption against the continuum that is red-shifted with respect to the $v_{\rm LSR}$ in the $^{13}$CO line is tracing a large scale infall, consistent with the findings at much lower resolution  of 5\arcsec\ ($\sim15000$\,au) by \citet{2013A&A...555A.112P}. 
This is also supported by the position velocity (pv) map of $^{13}$CO in Figure~\ref{fig:13co:pvmap}, where the morphology of the emission resembles the ``C'' shape characteristic of infall motions \citep{1997ApJ...488..241Z}.
H$_2$CO emission may be produced closer to the core center (or in the outflows) making harder the detection of infall signs (although the absorption dip is slightly red-shifted with respect to the $v_{\rm LSR}$). 

\subsubsection{Small Scale Expansion}

\begin{figure}
\plotone{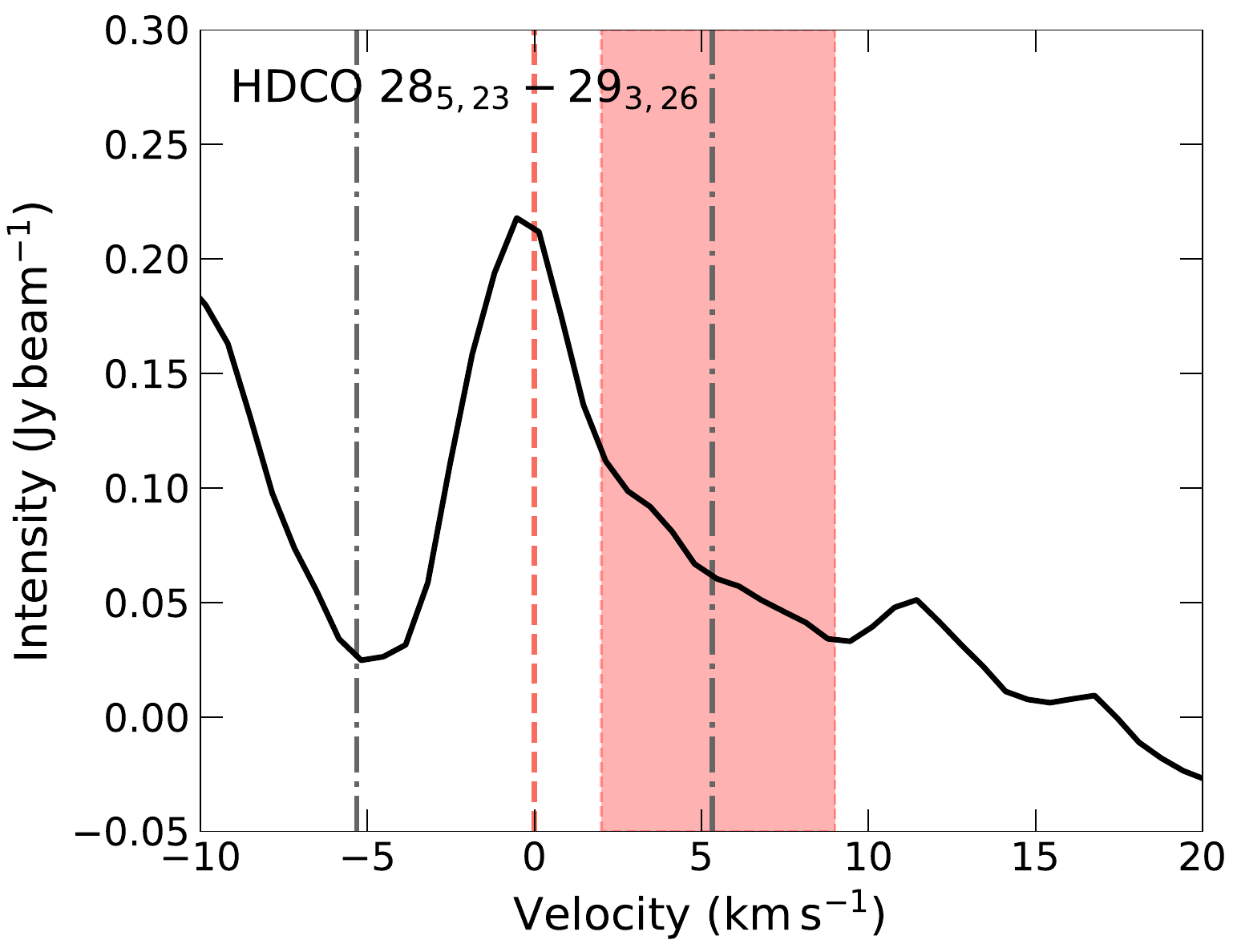}
\caption{HDCO $(28_{5,23} - 29_{3,26})$ line profile towards the continuum peak position G335--MM1 ALMA1. 
The dash-dotted lines show the limits used for the moments in Figure~\ref{fig:hotlines:moments}.
The red shadowed region highlights the red high-velocity wing.
\label{fig:hdco:line}}
\end{figure}

\begin{figure*}
\plotone{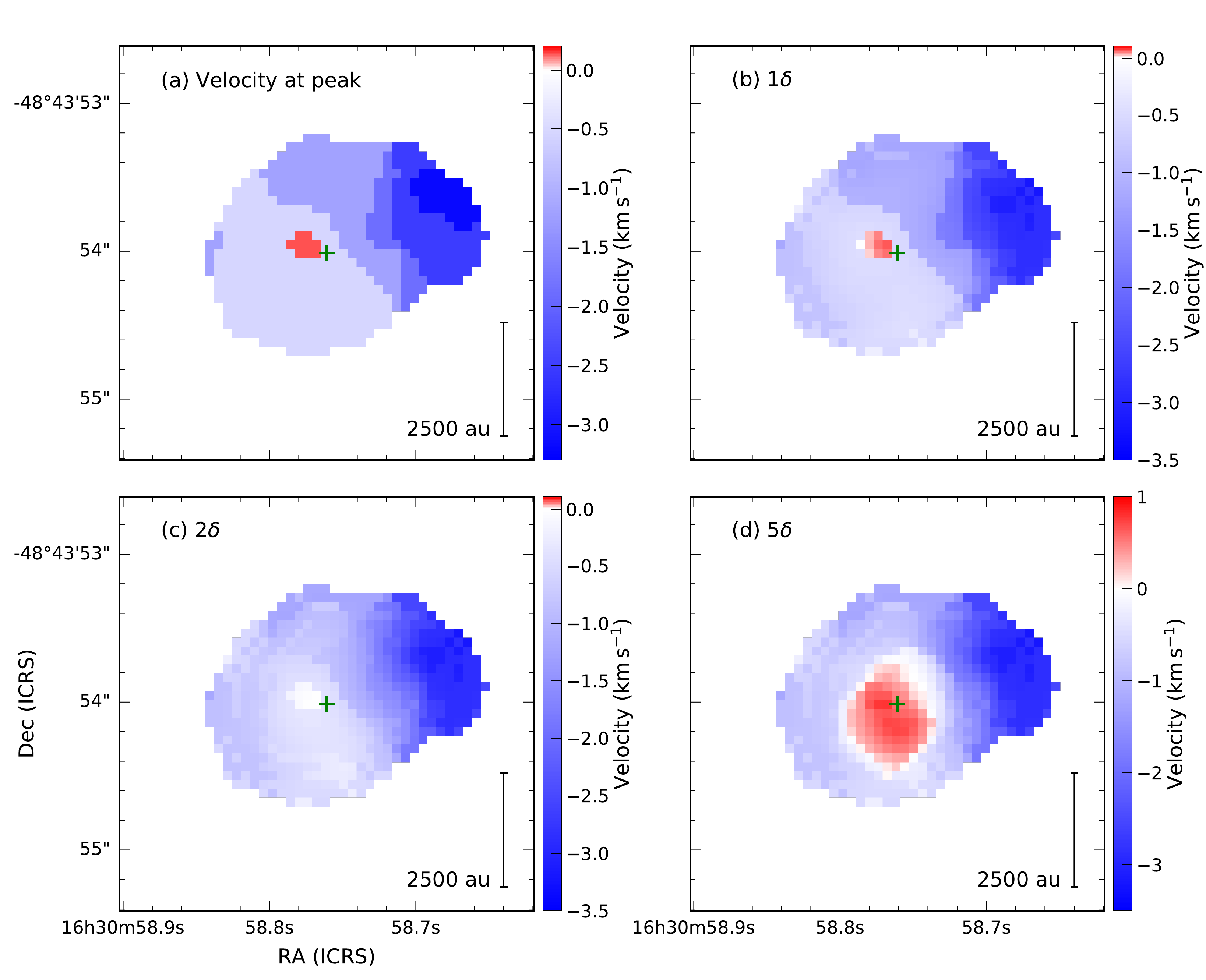}
\caption{HDCO $(28_{(5,23)} - 29_{(3,26)})$ velocity maps of G335--MM1 ALMA1. 
(a) Velocity at the line peak.
(b)--(d) First moment map integrated over $\pm1, 2, 5\times\delta$ from the line peak, respectively. 
The value of $\delta$ is defined as the median line standard deviation over the region, $\delta=2$ channels.
The green cross marks the position of the continuum source.
\label{fig:hdco:mom1}}
\end{figure*}

The spot at the core center in the first moment map (Figure~\ref{fig:hotlines:moments}) of the hot lines HDCO and (CH$_2$)$_2$CO is red-shifted with respect to the $v_{\rm LSR}$ (and not blue-shifted as in CH$_3$CN).
Moment maps have so far been made by using windows of $\pm5.2$ and $\pm6.5$\,\kms\ symmetric from the $v_{\rm LSR}$.
The HDCO line profile towards the continuum source position presented in Figure~\ref{fig:hdco:line} is single peaked with a red wing at higher velocities (shadowed region).
In order to discard the red wing as the cause of the red-shifted spot, we have derived the velocity structure in ALMA1 by first finding the peak at each pixel (defining a ``local" pixel $v_{\rm LSR}$) and then assuming different windows to make the first moment.
The velocity maps are shown in Figure~\ref{fig:hdco:mom1}.
This approach would be more robust to strong velocity gradients and we have no need to adopt large windows for making moment maps that can introduce noise or contamination from neighboring spectral lines. 
Figure~\ref{fig:hdco:mom1} shows that HDCO is red-shifted towards the center of the continuum source and increasingly blue-shifted further out, while towards the arc shaped structure the emission is blue-shifted as observed in CH$_3$CN. 
The area of the red-shifted region increases when more channels are included and shifts towards the south (cf. Figure~\ref{fig:hdco:mom1}c and d), indicating that the lines to the south are skewed towards the red due to high-velocity line wings.
Overall, the velocity gradient is consistent with what is seen in CH$_3$CN and as the window for making the moment map increases, the red-shifted spot becomes more evident.
Applying a similar reasoning for the origin of the blue-asymmetric/skewed lines (e.g.,  \citealp{1993ApJ...404..232Z}, see also the expansion line profiles in \citealp{2006ApJ...652.1366K}), this pattern likely indicates gas expansion at the core center.  
\citet{1987A&A...186..280A} noted that an inverted profile to that of blue-asymmetric profiles is expected for expansion motions if the temperature increases towards the central region and the velocities decrease outwards. 
The first condition is likely satisfied since a high-mass young stellar object may have already been formed and started to ionize the circumstellar gas.
They also assumed in their analysis that the Sobolev approximation is valid, but the expansion velocity of HDCO (see Section~\ref{sec-model}) is comparable to the line width.
Hence the second condition may not necessarily be satisfied.

\begin{figure}
\epsscale{1.15}
\plotone{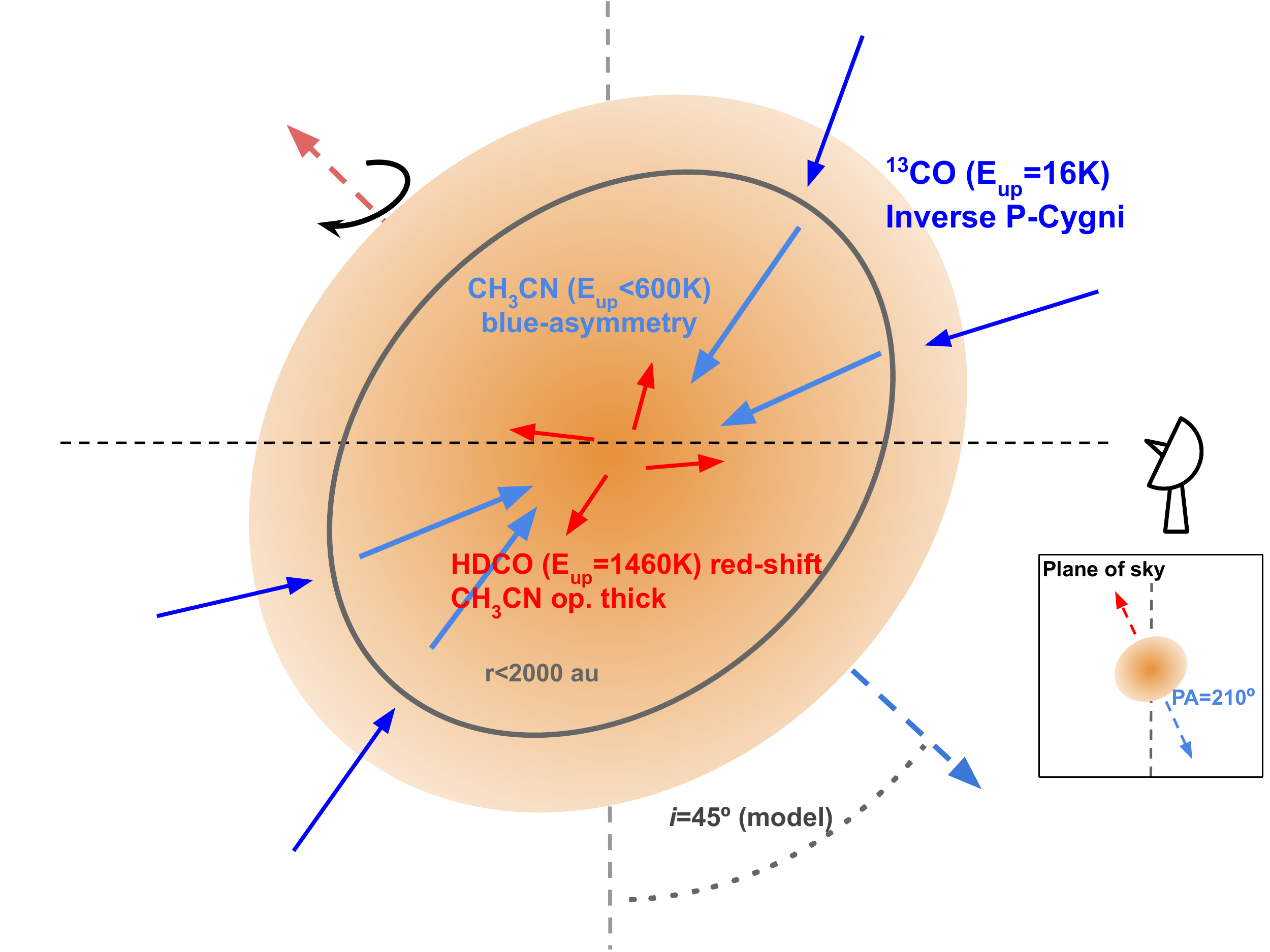}
\caption{Diagram of the motions within \regshort\ ALMA1 revealed by the observations and modeling.  \label{fig:alma1:diagram}}
\end{figure}

Using cm observations, \cite{2015A&A...577A..30A} determined that an HC \ion{H}{2} region, unresolved at their 1\farcs5 resolution observations, is located a the position of ALMA1. 
The HC \ion{H}{2} region is an important evolutionary stage in the life of high-mass stars. 
HC \ion{H}{2} regions tend to be smaller than 0.03\,pc (6000\,au) and the likely culprit of the ionized flow is the photo-evaporation of an accretion disk surface \citep[][]{2005IAUS..227..111K}.  
While at this stage, it has been suggested that the star can continue growing into earlier types by non-spherical accretion flows \citep{2007ApJ...666..976K}.
Eventually it is believed that HC \ion{H}{2} regions will expand into ultra compact (UC) \ion{H}{2} regions and then into classical \ion{H}{2} regions. 
We suggest that in ALMA1 we are witnessing the early expansion of the ionized gas that is pushing outward the hot molecular gas. 
The effect of the expansion is more clear in the transitions tracing the hot, inner molecular core (e.g., HDCO). 
The picture we propose for ALMA1 and its immediate surroundings is sketched in Figure~\ref{fig:alma1:diagram}.  
The spatial distribution of the emission from  the hot transition lines  (Figure~\ref{fig:hotlines:moments}a) is closer to perpendicular to the outflow (${\rm P.A.}=210\degr$), with position angles between 125\degr and 130\degr as measured from 2-D Gaussian fitted to the zeroth order moments. 
Hence, it is likely coming from molecular gas in a putative disk surface.

\subsection{ALMA3}

The $K=4$ transition first moment map in Figure~\ref{fig:ch3cn:k4b}a shows clear signs of rotation towards source ALMA3.
We estimate an average position angle of the rotation axis from the velocity gradient within a region of radius 0.16\arcsec\ (i.e., one beam) centered on the source of ${\rm P.A._{rot}}=292\pm8\degr$ from the first moment map of the $K=4$ transition.
A similar value is obtained from the $K=7$ transition but with a slightly higher error due to its smaller angular extent (${\rm P.A._{rot}}=290\pm9\degr$). 
We estimate the kinetic mass of source ALMA3 from the velocity extremes from the rotation axis.
Assuming that the source is edge-on, the source mass is in the 10--30~\msun\ range (hereafter kinetic mass).
Lower inclination angles, i.e., towards a face-on configuration, would imply an even larger mass.
This is one order of magnitude larger than the 1\,\msun\ derived from the dust emission (Table~\ref{tab:physprops}).
Note that the kinetic mass includes the contribution of the central (proto)star, which may not contribute to the dust emission, and the circumstellar gas.
On the other hand, the dust temperature estimation assumes an isothermal column of gas in LTE with the dust passing through the central (hotter) region. 
Given that the temperature can have a decreasing temperature profile with radius, the contribution of colder dust in the outer layers may be underestimated by these assumptions.
The temperature can reach values of roughly 30\,K at clump scales \citep[$>0.1$\,pc; e.g.,][]{2004A&A...426...97F}, hence a lower average temperature than our estimation could explain discrepancies of less than one order of magnitude (assuming dust emission in the Rayleigh-Jeans regime: $M_d \propto T_d^{-1}$).
Different dust opacity laws would explain discrepancies of a factor ${\sim}2$.
The lower number of lines detected and lack of radio emission suggest that this source is younger than ALMA1 and still deeply embedded.

\subsection{SIMPLE MODELING}
\label{sec-model}

In this section, we provide additional support for the interpretation of the moment maps by comparing qualitatively the data with simple LTE radiative transfer models. 
In these models, the molecular line emission arises from a spherically symmetric core with a radius of $a=2000$\,au, comparable to the extent of source ALMA1 shown in Figures~\ref{fig:ch3cn:k7a}, \ref{fig:hotlines:moments}, and \ref{fig:hdco:mom1}.
We assume that this core is characterized by a density $\rho(r) \propto r^{-3/2}$, and a thermal gradient with temperature $T(r)\propto r^{-0.5}$, characteristic of radiative equilibrium under optically thin dust conditions \citep{1985ApJ...296..655A}.
We calculated models with solid body rotation at each radius combined with infall or expansion motions.
The azimuthal velocity fields of the core vary as $V_\phi \propto r^{-1/2}$.
The radial velocities are proportional to $\pm r^{-1/2}$, with positive velocities for expansion motions and vice versa.
This model is a simplified version of a pressure-less free-falling core solution dominated by the gravity of a central object \citep{1976ApJ...210..377U,2004RMxAA..40..147M,2009MNRAS.393..579M}. 
Note that the models do not combine infall and expansion motions, thus they can only explain the features of one line at a time.
To fit the observations, we optimize numerically the central $v_{\rm LSR}$ and the line width. 
The remaining parameters are fine-tuned by visual inspection.

\begin{figure*}[b]
\plotone{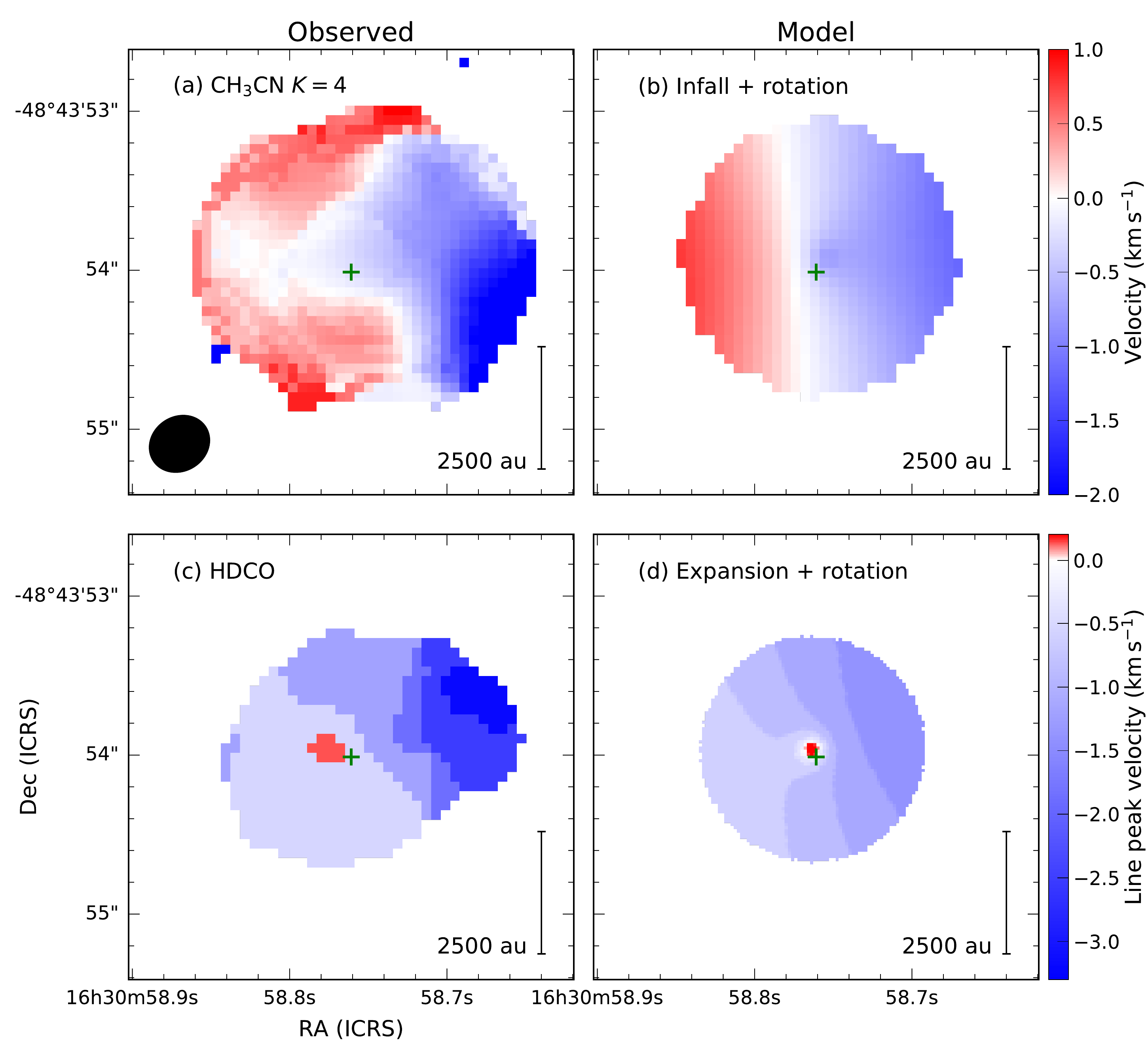}
\caption{Modeling of ALMA1 CH$_3$CN $J=12-11$ $K=4$ and HDCO observations.
(a) and (c) observed CH3CN $K=4$ moment 1 and line peak intensity of HDCO.
(b) Model of CH$_3$CN $K=4$ combining rotation and infall of a spherically symmetric core.
(d) Model of the line peak intensity of HDCO by a spherically symmetric core with expansion and rotation gas motions.
The beam is shown in the lower left corner of panel (a). \label{fig:models}}
\end{figure*}

The blue asymmetry, characteristic of infalling motions, requires a combination of partially optically thick emission and internal heating.
Figures~\ref{fig:models}a and \ref{fig:models}b shows the first moment map of the CH$_3$CN $J=12-11$ $K=4$ line towards ALMA1 and the one derived from the model. 
We assume that the rotation axis  of the core is inclined with respect to the line of sight by $i=45\degr$ and it forms a P.A.$=5\degr$, as indicated by the approximate direction of the velocity gradient. 
Note that the angular velocity of the core and the inclination angle are degenerate parameters.
We are able to reproduce the main features observed in ALMA1 with an infall velocity at the external radius $a$ of $V_{\rm in}(a)=-1.15$\,\kms, an angular velocity $\Omega(a)\sin(i)=3.34\times10^{-11}$\,s$^{-1}$, a temperature $T(a)=33$\,K, and a line absorption coefficient given by $\kappa_v(a)=3.21\times10^{-17} \phi_v$\,cm$^{-1}$. 
The line profile $\phi_v$ is assumed Gaussian with a FWHM of $\Delta v=2$\,\kms. 
We note that, in agreement with infall, the rotation of the core is not enough to maintain the core in equilibrium at the assumed inclination.

The gas mass within a radius $R$ from the model is given by
\begin{equation}\label{eq-mass-model}
    M(<R) = \frac{8\pi}{3} \rho(a) (Ra)^{3/2}\,.
\end{equation}
From the line absorption coefficient and assuming a CH$_3$CN abundance of $10^{-8}$ \citep[e.g.,][]{2014ApJ...786...38H}, we obtain $\rho(a)=2.9\times10^{-16}$\,g\,cm$^{-3}$.
The gas mass of the model at $R=710$\,au (same size derived from the dust continuum, see Table~\ref{tab:physprops}) is $M(<710\,{\rm au})=6.8$\,\msun.  
We note however that the abundance of CH$_3$CN is very uncertain and can vary by one order of magnitude \citep{2014ApJ...786...38H}, precluding a direct comparison with the mass derived from dust continuum emission.  

Figures~\ref{fig:models}c and \ref{fig:models}d show the maps of the HDCO velocity at peak line intensity obtained from the observations and a rotating and expanding core model.
The expansion velocity is \mbox{$V_{\rm exp}(a)=+0.5$\,\kms}.
Two features describe the shape of the HDCO line profile: 1) a relatively slow expansion which red-shifts the line peak in $\lesssim1$\,\kms\ in the directions of highest opacity, that is, toward the center of the core (see Figure~\ref{fig:hdco:mom1}a); 2) a distinct red-shifted wing, typical of outflows (see Figure~\ref{fig:hdco:line}).
A combination of these two effects produce the red-shifted velocities shown in the first moment map (Figure~\ref{fig:hdco:mom1}d).
We find that a model with an absorption coefficient 4 times less than that of CH$_3$CN and a line width $\Delta v=2.5$\,\kms\ is able to reproduce fairly well these features of the HDCO line, particularly the red-shifting of the peak toward the center of the core. 
This rotation-expansion model also has different P.A.$=20\degr$ compared to the CH$_3$CN model, which is suggested by the velocity gradient shown in Figure~\ref{fig:hdco:mom1}.
Models with the same P.A. produce fits that are not significantly worse.

As discussed in Section~\ref{sec:dis:alma1}, the expansion of the gas is likely caused by the expansion of the HC \ion{H}{2} (see Figure~\ref{fig:alma1:diagram}).
This may have already grown beyond the gravitational radius of the young star \citep{2019MNRAS.486.5171S}, which is 54\,au for a 10\,\msun\ HMYSO, at in the innermost regions of the hot core.
Note that  the lower opacity associated with the HDCO model implies that the lines of sight associated with $\tau=1$  cross the core much closer to its  center compared to  those of CH$_3$CN. 
Indeed, while for CH$_3$CN  the lines of sight with impact parameter of 130\,au are optically thick, for  HDCO, optically thick lines of sight are those that pass within 30\,au of the center. 
This difference in opacities explains why we are not able to see the expansion signature in the CH$_3$CN profiles and why we cannot discern the infall signature in the HDCO line. 
In the former case, the expansion is hidden within a very small radius associated with a large opacity. 
On the other hand, HDCO emission from gas in expansion is associated with optically thin emission, and therefore the blue asymmetry does not arise.

\section{Conclusions}\label{sec:conc}

We observed the high-mass star-forming region \region, in particular the core \regshort, at 0\farcs3 resolution with ALMA and resolved five sources within this region.
Of these sources, ALMA1 is associated with radio continuum emission previously observed at $\sim1-2$\arcsec\ resolution, while the other radio continuum source towards this region possibly arises from a jet associated with ALMA3.
From the study of the kinematics in these two sources, we conclude that they are likely to form or have already formed at least one high-mass star.
Line emission was not detected in the remaining three continuum sources, and were thus not studied in detail. 

ALMA1 has a complex kinematic structure.
We observe large scale infalling motions from $^{13}$CO inverse P-Cygni profiles, while CH$_3$CN blue asymmetric profiles indicate infall motions at smaller scales.
The overall CH$_3$CN velocity gradient may be indicative of rotation of the circumstellar material.
This velocity gradient is roughly perpendicular to the outflow direction observed from $^{13}$CO and SiO emission.
Finally, lines tracing hot molecular gas, HDCO and (CH$_3$)$_2$CO, show an expansion velocity pattern in their moment 1 map, which may be the result of photoevaporation of the surface of a molecular disk due to the ionizing radiation of the HC \ion{H}{2} region.
To support these hypotheses, we model the rotating infall and expansion motions with a spherically symmetric envelope.
We conclude that the expansion motion observed in hot lines is indicative of reversing of the accretion flow in a region smaller than the one traced by the rotating infall seen in CH$_3$CN.
Higher angular resolution observations will reveal the scales at which expansion dominates and whether or not the complex kinematics towards this source can also be the result of unresolved sources.

In ALMA3, CH$_3$CN line emission shows clear evidence of rotation, which imply a total mass for the source of 10--30\msun\ assuming Keplerian rotation.

The nature of the remaining sources (ALMA2, ALMA3, and ALMA5) remains unclear due to their lack of line emission. They likely are prestellar cores that may or may not form high-mass stars. 

\acknowledgments
The authors would like to thank the anonymous referee for the insightful comments.
F.O. and  H.-R.V.C. acknowledge the support of the Ministry of Science and Technology of Taiwan, project no. 109-2112-M-007-008-. 
P.S. was partially supported by a Grant-in-Aid for Scientific Research (KAKENHI Number 18H01259) of Japan Society for the Promotion of Science (JSPS). 
Data analysis was in part carried out on the Multi-wavelength Data Analysis System operated by the Astronomy Data Center (ADC), National Astronomical Observatory of Japan.
This paper makes use of the following ALMA data: ADS/JAO.ALMA\#2016.1.01036.S. ALMA is a partnership of ESO (representing its member states), NSF (USA) and NINS (Japan), together with NRC (Canada), MOST and ASIAA (Taiwan), and KASI (Republic of Korea), in cooperation with the Republic of Chile. The Joint ALMA Observatory is operated by ESO, AUI/NRAO and NAOJ.

\facility{ALMA}

\software{
astropy \citep{astropy:2013,astropy:2018},
CASA \citep{2007ASPC..376..127M},
GoContinuum \citep{olguin_fernando_2020_4302846},
matplotlib \citep{Hunter:2007},
numpy \citep{harris2020array},
scipy \citep{2020SciPy-NMeth},
YCLEAN \citep{2018ApJ...861...14C,2018zndo...1216881C}
          }
\newpage
\appendix

\section{Continuum subtraction}
\label{ap:contsub}

To obtain the line-free channels, we:
\begin{enumerate}
    \item Compute a dirty data cube for each spectral window in the observations.
    \item Compute a map with the maximum value along the spectral axis (hereafter maximum map) for each spectral window.
    \item Obtain the peak position of the maximum map for each spectral window.
    \item Average the values of the peak positions. 
    We rejected the position which is further from the centroid to avoid outliers.
    \item Obtain the spectrum at the averaged peak position from each spectral window data cube.
    \item Use asymmetric sigma clipping to obtain channels free of line contamination.
    \item Recover bands of channels rejected by the asymmetric sigma clipping that span less than two channels (one spectral resolution).
\end{enumerate}
In addition, 10 channels at each end of the spectral windows were not used for continuum calculations as those channels tend to be noisier.

A corrected symmetric sigma clipping has been implemented by \citet{2018A&A...609A.101S} to subtract the continuum from data cubes.
Their correction of the symmetric sigma clipping is based on the image noise, which cannot be calculated without having to CLEAN the data cube first.
Based on their approach, we found that an asymmetric sigma clipping can obtain similar results without applying their correction to the symmetric one.
Considering that our spectra are dominated by line emission we found that an acceptance range of $-3.0$ to $1.3\sigma_i$, with $\sigma_i$ the standard deviation at iteration $i$ of the clipping algorithm, is enough to obtain similar results than in \citet[][their default range is $\pm1.8\sigma_i$]{2018A&A...609A.101S}. What our method provides at the end is the list of channels, in CASA format, to be used for making both the continuum subtracted line cubes and the continuum image free of line contamination.

\bibliography{manuscript}{}
\bibliographystyle{aasjournal}



\end{document}